
\magnification\magstep1
\vsize23.0truecm 
\font\csc=cmcsc10
\font\bigbf=cmbx12 
\font\smallrm=cmr8 

\def\hatt{\widehat}
\def\half{\hbox{$1\over2$}}
\def\quart{\hbox{$1\over4$}}

\def\twentyfourth{\hbox{$1\over 24$}}
\def\d{{\rm d}}
\def\sumin{\sum_{i=1}^n}

\def\tilda{\widetilde}

\def\E{{\rm E}}
\def\RR{\mathord{I\kern-.3em R}}
\def\Var{{\rm Var}}
\def\arr{\rightarrow}

\def\section{\bigskip}
\def\subsection{\medskip}
\def\hop{\smallskip\noindent} 
\def\mise{\hbox{\csc mise}}
\def\amise{\hbox{\csc amise}}
\def\dna{\hbox{\csc dna}}
\def\ucv{\hbox{\csc ucv}}
\def\scv{\hbox{\csc scv}}
\def\Yepa{Yepanechnikov} 
\def\hhh{h_H}

\def\fermat#1{\setbox0=\vtop{\hsize4.00pc
        \smallrm\raggedright\noindent\baselineskip9pt
        \rightskip=0.5pc plus 1.5pc #1}\leavevmode
        \vadjust{\dimen0=\dp0
        \kern-\ht0\hbox{\kern-4.00pc\box0}\kern-\dimen0}}

\def\cstok#1{\leavevmode\thinspace\hbox{\vrule\vtop{\vbox{\hrule\kern1pt
        \hbox{\vphantom{\tt/}\thinspace{\tt#1}\thinspace}}
        \kern1pt\hrule}\vrule}\thinspace} 

\baselineskip14pt 

\bigskip
\centerline{\bigbf Towards Semiparametric Bandwidth Selectors}
\centerline{\bigbf for Kernel Density Estimators} 

\medskip
\centerline{\bigbf Nils Lid Hjort} 

\smallskip
\centerline{\bf University of Oslo} 

\smallskip
\centerline{\bf May 1999} 

\medskip
{{\smallskip\narrower\noindent 
\baselineskip12pt 
{\csc Abstract.} 
There is an intense and partly recent literature focussing 
on the problem of selecting the bandwidth parameter for kernel density
estimators. Available methods are largely `very nonparametric',
in the sense of not requiring any knowledge about the 
underlying density, or `very parametric', like the 
normality-based reference rule. This report aims at widening
the scope towards the inclusion of many semiparametric
bandwidth selectors, via Hermite type expansions around 
the normal distribution. The resulting bandwidths may
be seen as carrying out suitable corrections on the normal 
reference rule, requiring a low number of extra coefficients
to be estimated from data. 

The present report introduces and discusses some basic ideas 
and develops the necessary initial theory, but modestly 
chooses to stop short of giving precise recommendations 
for specific procedures among the many possible constructions. 
This will require some further analysis 
and some simulation-based exploration. 
Future work in this direction is planned with 
esteemed colleagues I.~Gijbels and M.C.~Jones. 

\smallskip\noindent\sl
{\csc Key words:} 
bandwidth selection, 
exact \mise,
integrated squared derivatives, 
Hermite approximations, 
kernel density estimation,
smoothed cross validation
\smallskip
}}

\section 
\centerline{\bf 1. Background, motivation and summary} 

\hop 
Suppose that $X_1,\ldots,X_n$ are independent 
and that it is required to estimate its density $f$. 
The classic kernel estimator is of the form 
$\hatt f(x)=n^{-1}\sumin K_h(X_i-x)$, 
in which $K_h(u)=h^{-1}K(h^{-1}u)$ is the $h$-scaled version
of $K$, the kernel function, which here is taken to
be a symmetric probability density function. 
This paper is about the already well-researched problem
of selecting a good bandwidth $h$ from data; 
consult for example Scott (1992), 
Wand and Jones (1995), Simonoff (1996),
Jones, Marron and Sheather (1996) or Chiu (1996) 
for recent and quite comprehensive overviews.

There is a strong tradition favouring mean integrated squared error
as performance criterion for density estimation.
The familiar decomposition into variance and squared bias gives 
$$\mise(h)=\E\int\{\hatt f(x)-f(x)\}^2\,\d x
        =\int\Var\hatt f(x)\,\d x + \int\{e_h(x)-f(x)\}^2\,\d x, \eqno(1.1)$$
where $e_h(x)$ is the mean of $\hatt f(x)$. 
Accordingly many efforts have been exuded on 
in some way or another estimating 
$$h_n=h_n(f), \quad {\rm the\ minimiser\ of\ }\mise(h), \eqno(1.2)$$
the best possible bandwidth value in this sense. 
\eject

Several of the best known approaches work with 
natural large-sample approximations to bias and variance.
These lead for example to 
$$\mise(h)\approx (nh)^{-1}R(K)+\quart k_2^2h^4R(f'')-n^{-1}R(f), \eqno(1.3)$$
where $k_2=\int u^2K(u)\,\d u$ while 
$R(p)=\int p(x)^2\,\d x$ is generic notation for functions $p$. 
One of the few methods that go directly for the exact \mise{}
is the so-called unbiased cross validation method that goes
back to Rudemo (1982) and Bowman (1984). 
It consists in minimising 
$$\ucv(h)=\int\hatt f(x)^2\,\d x
        -2n^{-1}\sumin \hatt f_{(i)}(X_i), \eqno(1.4)$$
where $\hatt f_{(i)}$ is the kernel estimator constructed from
the data set of size $n-1$ where $X_i$ is deleted;
note the dependence of $\hatt f$ and $\hatt f_{(i)}$ on $h$. 
The method derives its name from the fact that 
$\ucv(h)$ is an unbiased estimate of $\mise(h)$ minus the constant $R(f)$. 

The present paper is also concerned with methods that 
at least at the outset are strictly non-asymptotic 
in spirit, for estimating the exact $\mise$ curve and hence its minimiser. 
Its point of departure is a simple identity for $\mise$ 
entirely in terms of the density function $g$ 
for a difference of two data points. 
This identity, presented in Section 2,  
is neither complicated nor entirely new,
and is at least indirectly present in other literature,
see for example Kim, Park and Marron (1994, Section 2)
and Wand and Jones (1995, Ch.~2.6).
We are able to exploit this identity in several novel ways,
however, and in the process also shed some fresh light 
on some well-established $h$ selection methods,
including cross validation and the normal reference rule,
as well as smoothed cross validation and on traditional 
large-sample theory. This is discussed in Section 3. 

One conspicuous aspect of the identity is that only 
the $g$ density for $X_i-X_j$ matters, other aspects 
of the $f$ object are irrelevant. In particular $g$ is 
symmetric and often much more well-behaved than the original $f$,
and consequently more amenable to statistical modelling. 
This viewpoint invites several natural semiparametric
$h$ selection rules. These consist in minimising an
estimate of the \mise{} curve constructed by insertion 
of a suitable semiparametric density estimate for $g$. 
This is illustrated in Section 4 for a family 
of Hermite expansion based models. 
These $h$ rules are easy to use and can be seen as 
`extended rules of thumb' that in various ways correct for 
non-normal features of $g$. 
In practice these rules will typically be less variable 
than most of the purely nonparametric ones, but will
perhaps aim at a slightly sub-optimal $h$.
Such a rule might nevertheless often win over sophisticated 
nonparametric ones, in that a small bias and smaller variance
often beats zero bias and bigger variance. 
Theory is developed in Section 5, with clarification 
of the basic bias and variance issues involved in the process. 
An initial and not yet conclusive discussion is given in Section 6.
Finally some extraneous results and comments are offered 
in Section 7. 

In bandwidth selection literature the big space between
parametric and nonparametric methods has been left 
quite un-explored. This paper can be viewed as reporting on 
initial discoveries and consequent problems related to 
attempts of opening up this box of possibilities. 
It is admitted that many of the questions emerging from the box, 
partly due to the large number of sensible new constructions, 
need further attention before they can be conclusively answered.  
In particular, further work and simulation explorations are 
needed to determine which of the indicated proposals of 
Section 6 are more fruitful than others. 

\section 
\centerline{\bf 2. The exact MISE}

\hop 
The aim of this section is to work out a fruitful
exact expression for \mise{} in terms of 
$$g(y)=\int f(x)f(x+y)\,\d x, \eqno(2.1)$$
the density of a difference $Y_{i,j}=X_j-X_i$ between
two different data points. Let similarly 
$g_K(y)=\int K(u)K(u+y)\,\d u$ for the kernel function $K$. 

{\smallskip
{\csc Proposition.} \sl 
The exact \mise{} for the kernel estimator can be expressed as 
$$\eqalign{\mise(h)
&=(nh)^{-1}R(K)+\int A_K(v)g(hv)\,\d v+R(f) \cr
&=(nh)^{-1}R(K)+\int h^{-1}A_K(h^{-1}y)g(y)\,\d y+R(f), \cr} \eqno(2.2)$$
where $A_K(v)=(1-n^{-1})g_K(v)-2K(v)$.
\smallskip}

To prove this, note first that the mean of 
$\hatt f(x)$ can be expressed as 
$$e_h(x)=\E K_h(X_i-x)=
        \int K_h(y-x)f(y)\,\d y=\int K(u)f(x+hu)\,\d u. $$ 
Similarly its variance can be written
$(nh)^{-1}a_h(x)-n^{-1}e_h(x)^2$, 
in which $a_h(x)=\int K(u)^2\allowbreak f(x+hu)\,\d u$. 
Next observe that $\int a_h(x)\,d x=\int K(u)^2\,\d u=R(K)$,
after interchanging order of integrals,
so that the integrated variance becomes equal to 
$(nh)^{-1}R(K)-n^{-1}\int e_h^2\,\d x$;
thus there is nothing asymptotic about the familiar $(nh)^{-1}R(K)$ 
term that starts out the \mise{} approximations of type (1.3). 
Next two similar Fubini operations lead to 
$\int e_h^2\,\d x=\int g_K(v)g(hv)\,\d v$. 
Also, $\int e_hf\,\d x$ is seen to be the same as $\int K(v)g(hv)\,\d v$. 
This gives the required result via (1.1). 
\smallskip

We note that the $A_K$ function is almost independent of $n$,
and that the $R(f)=g(0)$ term of course is irrelevant for
the minimisation of \mise. In other words, 
the direct, non-asymptotic function to minimise is 
$$\dna(h)=(nh)^{-1}R(K)+q(h)=\mise(h)-R(f), \eqno(2.3)$$
where 
$$\eqalign{q(h)
&=\E\,h^{-1}A_K(h^{-1}Y) \cr
&=(1-n^{-1})\int h^{-1}g_K(h^{-1}y)g(y)\,\d y
        -2\int h^{-1}K(h^{-1}y)g(y)\,\d y, \cr} \eqno(2.4)$$
writing $Y$ for a generic difference $X_j-X_i$ 
between two different data points. 
Again, note that $q(h)$ is almost independent of $n$. 

We shall take special interest in two popular kernel
functions. The first is the the standard normal $K=\phi$,
for which $g_K$ is the $N(0,2)$ density. 
The second is the one that manages to minimise both the pointwise 
and integrated mean squared error for large samples, 
namely the \Yepa{} kernel 
$K(u)={3\over2}(1-4u^2)$ for $|u|\le\half$. 
For this kernel function somewhat long calculations give 
$$g_K(y)=\cases{(6/5)(1-5y^2+5y^3-y^5) &for $0\le y\le 1$, \cr
        (6/5)(1-5y^2-5y^3+y^5)         &for $-1\le y\le 0$, \cr}$$
see Byholt and Hjort (1999). 

\section 
\centerline{\bf 3. Normal reference, UCV and SCV rules}
\centerline{\bf and asymptotic approximations revisited}

\hop  
In view of the result above a group of natural 
$h$ selection procedures emerges:
for each well-motivated proposal $\hatt q(h)$ for estimating $q(h)$, form 
$$\hatt{\dna}(h)=(nh)^{-1}R(K)+\hatt q(h), \eqno(3.1)$$
and let $\hatt h$ be its minimiser. 
This and the following section consider several appealing special cases.
The present section in particular sheds 
some fresh light on several well-established bandwidth 
selection strategies, while several new rules 
emerge naturally in the following sections. 

\subsection 
{\sl 3.1. Normal difference density: the normal reference rule in new light.}
At one extreme is the hard parametric assumption that 
$Y$ is normal. Let us write this as $Y\sim N(0,2\sigma^2)$,
in terms of the standard deviation $\sigma$ for $X_i$
($Y$ is necessarily symmetric about zero). 
With $g(u)=(\sqrt{2}\sigma)^{-1}\phi((\sqrt{2}\sigma)^{-1}u)$ 
function (2.2) can be minimised over $h$. The result is of the form
$h_n=b_n\sigma/n^{1/5}$, where $b_n$ is the well-defined 
minimiser of the function $\mise_0(b/n^{1/5})$.  
Here $\mise_0(h)$ is as in (2.2), calculated under 
$g=N(0,2)$, that is, with $\sigma=1$. 
This leads to the proposal 
$$\hatt h_n=b_n\hatt\sigma/n^{1/5}, \eqno(3.2)$$
inserting any reasonable estimator for $\sigma$. 

With a standard normal kernel one finds from (2.4) that 
$q(h)=\sigma^{-1}q_0(\sigma^{-1}h)$, where 
$q_0(\cdot)$ is computed under $\sigma=1$, giving 
$$q_0(h)=(2\sqrt{\pi})^{-1}\{(1-n^{-1})(1+h^2)^{-1/2}
        -2(1+\half h^2)^{-1/2}\}. $$
The method is as in (3.2) with $b_n$ chosen to minimise 
$$\mise_0(b/n^{1/5})={1\over 2\sqrt{\pi}}
        \Bigl\{{1\over n^{4/5}b}+\Bigl(1-{1\over n}\Bigr)
        {1\over (1+b^2/n^{2/5})^{1/2}}
        -2{1\over (1+\half b^2/n^{2/5})^{1/2}}\Bigr\}. $$
This is a more precise finite-sample version of the 
classical $(4/3)^{1/5}\hatt\sigma/n^{1/5}$ rule
which stems from minimising the approximate \mise{} as in (1.3)
under a normal reference assumption for $f$,
cf.~Silverman (1986, Ch.~3.4) and Scott (1992, Ch.~6.2). 
The table below illustrates that with finite $n$ one
should stretch $h$ a little more, to gain a couple of
percent in \mise{} reduction. 

For the \Yepa{} kernel similar but much longer calculations
are provided in Byholt and Hjort (1999), with the 
$g_K$ function found above as one ingredient. 
Again the result is an $h_n$ of the form $c_n\sigma/n^{1/5}$,
where $c_n$ minimises a well-defined function of $c$ 
for the given $n$, and the normal reference rule becomes 
$\hatt h_n=c_n\hatt\sigma/n^{1/5}$. 
The simplistic version of this, using (1.3) 
with a normal reference for $f$, is $4.6898\,\hatt\sigma/n^{1/5}$. 
A small table of $b_n$ and $c_n$ is given here.

{{\medskip\narrower\narrower\baselineskip11pt\noindent
{\csc Table 1.} 
{\sl Optimal reference rules for the normal and for
the \Yepa{} kernels: these are respectively 
$b_n\hatt\sigma/n^{1/5}$ and $c_n\hatt\sigma/n^{1/5}$,
where $b_n$ and $c_n$ are given here for some sample sizes.} 

\smallskip
\baselineskip10pt\parindent0pt
\obeylines\tt
~~3~~1.2871~~5.2821~\qquad~~~~~~~~14~~1.1849~~5.0198
~~4~~1.2628~~5.2177~\qquad~~~~~~~~15~~1.1816~~5.0117
~~5~~1.2458~~5.1737~\qquad~~~~~~~~16~~1.1786~~5.0043
~~6~~1.2331~~5.1411~\qquad~~~~~~~~17~~1.1759~~4.9975
~~7~~1.2230~~5.1156~\qquad~~~~~~~~18~~1.1734~~4.9913
~~8~~1.2148~~5.0949~\qquad~~~~~~~~19~~1.1711~~4.9855
~~9~~1.2080~~5.0776~\qquad~~~~~~~~20~~1.1689~~4.9801
~10~~1.2021~~5.0628~\qquad~~~~~~~~50~~1.1368~~4.8996
~11~~1.1970~~5.0500~\qquad~~~~~~~100~~1.1190~~4.8540
~12~~1.1925~~5.0388~\qquad~~~~~~1000~~1.0842~~4.7617
~13~~1.1885~~5.0288~\qquad~~~~~~~~$\infty$~~1.0592~~4.6898
\medskip}}

\subsection 
{\sl 3.2. Why does the normal reference rule work so well?} 
Note that assuming a normal density for $Y=X_1-X_2$ 
is not only implied by but actually also 
equivalent to having a normal density for $X$. 
This is a non-trivial mathematical characterisation theorem
for the normal distribution. 
However, even though the two statements `$f$ is normal'
and `$g$ is normal' are mathematically equivalent,
it is fair to say that $g$ is often much more normal than $f$.
The $f$ to $g$ operation averages and symmetrises at the same time,
and the result is smoother and indeed `more normal'. 
One way of appreciating closeness to normality is to look at 
closeness to zero of all cumulants from order three on.
Here the odd ones vanish while the even ones decrease; 
the $2j$th cumulant coefficient for $g$ is equal to 
$(\half)^{j-1}$ times that of $f$, for $j\ge2$. 

\subsection
{\sl 3.3. Direct nonparametric: the \ucv{} revisited.}
The simplest straightforward nonparametric proposal 
estimates the theoretical mean by the empirical mean 
over all differences, giving
$$\hatt{\dna}(h)=(nh)^{-1}R(K)+{1\over n(n-1)} 
        \sum_{i\not=j}h^{-1}A_K(h^{-1}Y_{i,j}); \eqno(3.3)$$
the $A_K$ function is symmetric so we only have 
to take the mean of the summands with $i<j$ in practice.
This is the natural assumption-free unbiased estimate of 
$\dna(h)$. But it turns out via some algebraic manipulations 
that this is exactly the same as the \ucv{} formula (1.4). 

One might argue that (3.3) has an even more intuitive 
motivation than the the \ucv{} in the form (1.4).
In particular the unbiasedness property,
which with the (1.4) approach takes a little bit of 
algebra to demonstrate, is very direct here. 

\subsection 
{\sl 3.4. From estimation of g to smoothed cross validation.}
There is also an interesting link from the present 
approach, estimating $\dna$ curves, to the so-called 
smoothed cross validation method, as studied by 
M\"uller (1985), Staniswalis (1989), Hall, Marron and Park (1992)
and Wand and Jones (1995, Section 3.7), among others. 
This link will not be discussed further here,
but rather in future work with Ir\`ene Gijbels and Chris Jones. 

\subsection
{\sl 3.5. Links to large-sample theory.}  
Here we indicate how the familiar approximations to \mise{} 
of the type (1.3) can be derived via our $g$-based identity (2.2).
Some finite-sample corrections to these are also suggested
by the following simple calculations. 
This will also help us later in understanding the 
behaviour of the Hermite selectors. 

The trick is to start with the first formula in (2.2),
and then Taylor expanding $g(hv)$ for small $h$. 
Hence the $q(h)=\int A_K(v)g(hv)\,\d v$ term can be approximated with 
$$\eqalign{
g(0)\{&(1-n^{-1})\cdot1-2\}
+\half h^2g''(0)\{(1-n^{-1})2k_2-2k_2\} \cr
&+\twentyfourth h^4g^{(4)}(0)\{(1-n^{-1})(2k_4+6k_2^2)-2k_4\} \cr 
&+\hbox{${1\over 720}$} h^6g^{(6)}(0)\{(1-n^{-1})(2k_6+30k_2k_4)-2k_6\}
        +\cdots, \cr}$$
using $k_j$ for the $j$th moment of $K$. 
Thus 
$$\eqalign{
q(h)&\doteq -(1+n^{-1})g(0)+\quart k_2^2h^4g^{(4)}(0)(1-n^{-1}) \cr 
&\qquad\qquad 
        +\twentyfourth k_2k_4h^6g^{(6)}(0)(1-n^{-1})-k_2h^2n^{-1}g''(0), \cr}$$
omitting terms of lesser importance. 
Note now that in addition to $g(0)=R(f)$ we have 
$g''(0)=\int ff''=-R(f')$, $g^{(4)}(0)=\int ff^{(4)}=R(f'')$,
and so on, under sufficient regularity assumptions on the density.
In particular, using the (2.2) identity again, 
we recognise the familiar (1.3) approximation. 
It also follows from the above 
that the optimal bandwidth $h_n$ of (1.2) 
for moderate to large $n$ can be approximated with the 
explicit minimiser, say $h_{n,a}$, of the asymptotic \mise{} 
expression (1.3), in that 
$$h_n=h_{n,a}+O(n^{-3/5})
  =\{R(K)/k_2^2\}^{1/5}\{g^{(4)}(0)\}^{-1/5}n^{-1/5}+O(n^{-3/5}). \eqno(3.4)$$
See for example Wand and Jones (1995, Ch.~3) or Fan and Marron (1992). 
\eject 

Several of our $h$ selection schemes rely on estimating 
$q(h)$ by plugging in a nonparametric or partly parametric 
estimate $\hatt g(y)$ for $g(y)$. An exact parallel to 
the calculation above shows that in that case, 
$$\hatt q(h)=-(1+n^{-1})\hatt g(0)+\quart k_2^2h^4\hatt g^{(4)}(0)(1-n^{-1}) 
        +\twentyfourth k_2k_4h^6\hatt g^{(6)}(0)(1-n^{-1}) $$
plus terms that are either smaller in size or independent of $h$. 
Several matters become clearer in this light. 

Note first that we do not really need an ambitiously
complete model for the $g$ density;
what matters is reasonable estimation of the quantity 
$\int A_K(v)g(hv)\,\d v$ for $h$ of size $n^{-1/5}$.
The crux, at least for large $n$, is that 
the fourth derivative of $\hatt g$ of zero should be close
to the real $g^{(4)}(0)=R(f'')$ with high probability. 
Next note that this sometimes can be achieved by using 
a partly or fully parametric model, say $g(y,\theta)$,
where there could be modelling bias but smaller sampling
variability in comparison with the fully nonparametric schemes. 
For large $n$ the modelling bias would be measured by the ratio 
$\rho=\{g^{(4)}(0,\theta_0)\}^{-1/5}/\{g^{(4)}(0)\}^{-1/5}$,
where $\theta_0$ is the limit in probability of $\hatt\theta$. 
The variance of $\hatt h_n/h_n$ would be of size $O(n^{-1})$
under traditional parametric assumptions, and this is
smaller than for many of the nonparametric $h$ rules. 
Thus $h$ rules of this type are good whenever 
the $\rho$ ratio is close to 1 
and the number of parameters used is small. 
For smaller $n$ the more precise quantity to consider 
would be $\rho_n=h_{n,\theta_0}/h_n$, 
where $h_n$ is the exact \mise-minimiser and 
$h_{n,\theta}$ minimises the appropriate $(nh)^{-1}R(K)+q(h,\theta)$. 

The following section introduces extended $h$ rules
via Hermite extensions. How successful can they be? 
In view of remarks above this is essentially determined 
by two issues, at least for large $n$.  
The first is closeness of the approximation of the model's 
$g^{(4)}(0)^{1/5}$, say $R_{2m}^{1/5}$ in an expansion
with $2m$ terms, to the real quantity $R(f'')^{1/5}$.
The second issue is that of small enough sampling variability 
of the estimate of this quantity, say $\hatt R_{2m}^{1/5}$. 
These issues are investigated in Section 5. 

\section 
\centerline{\bf 4. New rules using short Hermite expansions}

\hop
This section considers a new class of natural selection rules 
based on the (3.1) minimisation recipe, through
models for the symmetric $g$ density that are 
far less restricted than the normal but also 
not as wide than the utterly nonparametric.  
The tools will be those of approximations to
densities of the form a normal times short Hermite polynomial
expansions, which we review first. 

\subsection
{\sl 4.1. Approximations using Hermite expansions.}
Let $H_j$ be the $j$th Hermite polynomial, defined via
$\phi^{(j)}(x)=(-1)^j\phi(x)H_j(x)$; the first few are
$H_0=1$, $H_1=x$, $H_2=x^2-1$, $H_3=x^3-3x$, and
$H_4=x^4-6x^2+3$. These have the property that 
$\int H_j^2\phi\,\d x=j!$ while 
$\int H_jH_k\phi\,\d x=0$ for $j\not=k$. 

Now consider in general terms the possibility of 
approximating a given density $p(x)$ with a standard normal 
times an additive expansion. For any positive $\hhh$ 
the functions $H_j(\hhh^{-1}x)$ are orthogonal with respect to 
$\hhh^{-1}\phi(\hhh^{-1}x)$. To form an expansion, minimise 
$$\eqalign{
\int\Bigl\{&{p(x)\over \phi(x)}
        -\sum_{j=0}^m a_jH_j(\hhh^{-1}x)\Bigr\}^2
        \hhh^{-1}\phi(\hhh^{-1}x)\,\d x \cr 
&\qquad\qquad 
  =\int\Bigl\{p(x)-\phi(x)\sum_{j=0}^m a_jH_j(\hhh^{-1}x)\Bigr\}^2
        {\hhh^{-1}\phi(\hhh^{-1}x)\over \phi(x)^2}\,\d x \cr}$$
with respect to the coefficients $a_0,\ldots,a_m$. 
This brings forwards the best approximation of order $m$, 
for given $\hhh$, namely 
$$p_m(x)=\phi(x)\sum_{j=0}^m{\alpha_j\over j!}H_j(\hhh^{-1}x)
        \quad {\rm with} \quad 
\alpha_j=\E_p H_j(\hhh^{-1} X)
        {\hhh^{-1}\phi(\hhh^{-1}X)\over \phi(X)}. \eqno(4.1)$$

In our applications we shall have occasion 
to consider several possible values of $\hhh$. 
The standard choice is actually $\hhh=1$, 
leading to the direct Hermite expansion 
$p_m(x)=\phi(x)\sum_{j=0}^m(\gamma_j/j!)\,H_j(x)$
which uses $\gamma_j=\E H_j(X)$. 
For a standardised $X$ with mean zero and variance one this 
gives $\gamma_0=1$, $\gamma_1=\gamma_2=0$, and then the familiar 
skewness, kurtosis and so on for $\gamma_3$, $\gamma_4$ and so on.
While well-known and possibly popular,
also due to its connection to cumulants,  
it is easy to see that this direct expansion has serious drawbacks; 
the moments of $p$ must be finite and the natural 
sample estimates become quite noisy and non-robust. 
There are also cases in which all coefficients exist but 
where the expansion simply does not converge; see Fenstad and Hjort (1997). 
Part of the reason is visible from the arguments above;
this expansion corresponds to using weighted $L_2$ 
as criterion function, with weight function $1/\phi(x)$,
or $\exp(\half x^2)$, which means extreme weight for 
rather non-central $x$ values. 
A more natural choice, in this light, is $\hhh=1/\sqrt{2}$, 
corresponding to canonical, non-weighted $L_2$ distance. 
The result is what is called the robust Hermite expansion
in Fenstad and Hjort (1997), 
$p_m(x)=\phi(x)\sum_{j=0}^m(\delta_j/j!)\,H_j(\sqrt{2}x)$
with $\delta_j=\sqrt{2}\E_pH_j(\sqrt{2}X)\exp(-\half X^2)$. 

However, in the situation at hand we are interested in 
applying this machinery to approximating and then 
estimating the difference density $g(y)$ with a view towards 
achieving best precision around zero, 
cf.~the $\int A_K(v)g(hv)\,\d v$ formula for $q(h)$. 
This wish could be reflected in choosing
a weight function for the $L_2$ criterion, 
whose form in the setting above is $\exp\{-\half(1/\hhh^2-2)x^2\}$,
with most weight around zero. This corresponds to 
choosing a smaller value for $\hhh$. That this is a good idea 
will be elaborated on in Section 5. 
There is a balance to be struck in that too small 
$\hhh$ values leads to higher estimation variability. 

\subsection 
{\sl 4.2. Expressions for MISE and its estimation.}
Since the $g$ density is symmetric with variance $2\sigma^2$, 
the theory above naturally lends itself 
to the approximation of the density for $Y/(\sqrt{2}\sigma)$ 
around the standard normal. 
Hence we study the expansion model 
$$g_{2m}(y)=\phi\Bigl({y\over \sqrt{2}\sigma}\Bigr){1\over \sqrt{2}\sigma}
        \sum_{j=0}^m{\alpha_{2j}\over (2j)!}
        H_{2j}\Bigl({\hhh^{-1}y\over \sqrt{2}\sigma}\Bigr). \eqno(4.2)$$
The odd moments vanish by symmetry, and 
$$\eqalign{\alpha_{2j}
&=\E\,H_{2j}\Bigl({\hhh^{-1}Y\over \sqrt{2}\sigma}\Bigr)
        {\hhh^{-1}\phi(\hhh^{-1}Y/(\sqrt{2}\sigma))\over 
        \phi(Y/(\sqrt{2}\sigma))} \cr 
&=\int \hhh^{-1} H_{2j}\Bigl({\hhh^{-1} y\over \sqrt{2}\sigma}\Bigr)
        \exp\Bigl\{-\half(1-\hhh^2)
        \Bigl({\hhh^{-1}y\over \sqrt{2}\sigma}\Bigr)^2
                \Bigr\}g(y)\,\d y. \cr}\eqno(4.3)$$
One way of estimating the $\alpha_{2j}$s is presumably 
by first deriving expressions for these $Y$-related 
quantities in terms of related $X$-related ones.
This seems difficult in any generality, however, 
and it is simpler to bypass the $X_i$s and use the $Y_{i,j}$s
directly. We use
$$\hatt\alpha_{2j}={n\choose2}^{-1}\sum_{i<l}
        \hhh^{-1} H_{2j}\Bigl({\hhh^{-1}Y_{i,l}\over \sqrt{2}\sigma}\Bigr)
        \exp\Bigl\{-\half(1-\hhh^2)
\Bigl({\hhh^{-1}Y_{i,l}\over \sqrt{2}\sigma}\Bigr)^2\Bigr\} \eqno(4.4)$$
(noting that all $H_{2j}$ functions are symmetric). 
One would typically insert an estimate $\hatt\sigma$ for $\sigma$ here,
but it is useful to note that the expansion methods as such
allow any positive value to be used here. 

We now provide a formula for $q(h)$ of (2.4) in the case 
of a normal kernel. It is helpful to write
this as $\sigma^{-1}q_0(\sigma^{-1}h)$ again, 
where it suffices to find the $q_0(\cdot)$ 
that comes from putting $\sigma=1$. One finds 
$q_0(h)$ of the form $(1-n^{-1})S_0(h)-2T_0(h)$, say. 
To find these terms, let $1/a=(1+1/h^2)^{1/2}$
and $1/b=(\half+1/h^2)^{1/2}$, that is, 
$a=h/(1+h^2)^{1/2}$ and $b=h/(1+\half h^2)^{1/2}$. 
Then 
$$\eqalign{S_0(h)
&=\int\phi\Bigl({y\over \sqrt{2}h}\Bigr){1\over \sqrt{2}h}
        \phi\Bigl({y\over \sqrt{2}}\Bigr){1\over \sqrt{2}}
         \sum_{j=0}^m{\alpha_{2j}\over (2j)!}
        H_{2j}\Bigl({\hhh^{-1} y\over \sqrt{2}}\Bigr)\,\d y \cr
&={1\over 2\sqrt{\pi}}{a\over h}
        \sum_{j=0}^m{\alpha_{2j}\over (2j)!}\E_N(\hhh^{-1}a Z) \cr} $$
and similarly 
$$\eqalign{T_0(h)
&=\int\phi\Bigl({y\over h}\Bigr){1\over h}
        \phi\Bigl({y\over \sqrt{2}}\Bigr){1\over \sqrt{2}}
        \sum_{j=0}^m{\alpha_{2j}\over (2j)!}
        H_{2j}\Bigl({\hhh^{-1} y\over \sqrt{2}}\Bigr)\,\d y \cr
&={1\over 2\sqrt{\pi}}{b\over h}
        \sum_{j=0}^m{\alpha_{2j}\over (2j)!}
        \E_NH_{2j}(\hhh^{-1}b Z/\sqrt{2}), \cr}$$
where the expectations are with respect to a $Z$ being standard normal. 
These formulae can be worked with further through first 
detecting and then proving that 
$\E_NH_{2j}(cZ)=1\cdot3\cdots(2j-1)(c^2-1)^j$,
which also can be written $\{(2j)!/(2^j j!)\}\,(c^2-1)^j$. 
Thus one finds 
$$\eqalign{
S_0(h)&={1\over 2\sqrt{\pi}}{1\over (1+h^2)^{1/2}}
        \Bigl[\alpha_0+\sum_{j=1}^m{\alpha_{2j}\over 2^jj!}(-1)^j
        \Bigl\{{1-h^2(1-\hhh^2)/\hhh^2\over 1+h^2}\Bigr\}^j\Bigr], \cr
T_0(h)&={1\over 2\sqrt{\pi}}{1\over (1+\half h^2)^{1/2}}
        \Bigl[\alpha_0+\sum_{j=1}^m{\alpha_{2j}\over 2^jj!}(-1)^j
        \Bigl\{{1-\half h^2(1-\hhh^2)/\hhh^2
                \over 1+\half h^2}\Bigr\}^j\Bigr]. \cr}$$
There are simplifications for the Hermite expansion 
standard choice $\hhh=1$, but we shall typically want to use 
smaller values. Recall here that the $\alpha_{2j}$s depend on $\hhh$. 

Observe further that normality corresponds to all $\alpha_{2j}$s 
being equal to zero except for $\alpha_0=1$, 
in which case we get back the formula used in Section 3.1.
One natural proposal is therefore to use bandwidth $\hatt h$, 
depending on both $\hatt\sigma$ and the $\hatt\alpha_{2j}$s 
that have been chosen for inclusion, 
defined as the minimiser of the easily programmed function 
$$\eqalign{\hatt{\dna}(h)
&=(2\sqrt{\pi})^{-1}(nh)^{-1}
        +(1-n^{-1})\,\hatt\sigma^{-1}\hatt S_0(\hatt\sigma^{-1}h)
        -2\,\hatt\sigma^{-1}\hatt T_0(\hatt\sigma^{-1}h) \cr 
&={1\over 2\sqrt{\pi}}\Bigl[{1\over nh}
        +{1-n^{-1}\over (\hatt\sigma^2+h^2)^{1/2}}
        \sum_{j=0}^m {\hatt\alpha_{2j}\over 2^j j!}(-1)^j
        \Bigl\{{\hatt\sigma^2-h^2(1-\hhh^2)/\hhh^2
        \over \hatt\sigma^2+h^2}\Bigr\}^j \cr 
&\qquad\qquad 
        -2{1\over (\hatt\sigma^2+\half h^2)^{1/2}}
        \sum_{j=0}^m{\hatt\alpha_{2j}\over 2^j j!}(-1)^j
        \Bigl\{{\hatt\sigma^2-\half h^2(1-\hhh^2)/\hhh^2
        \over \hatt\sigma^2+\half h^2}\Bigr\}^j\Bigr]. \cr}\eqno(4.5)$$
Here $\hatt S_0(\cdot)$ and $\hatt T_0(\cdot)$ are as above but 
with estimates inserted for the $\alpha_{2j}$s included. 

\smallskip
{\csc Remark 1.} 
The $h$ selection scheme laid out here requires the prior
selection of two pilot parameters, 
the expansion order $m$ and the Hermite approximation bandwidth $\hhh$.
Some advice on this will be implicit in results of 
the next section, and a more careful study will take 
place in future work. The next section also discusses 
certain modifications that lead to better performance than that 
achievable using the direct method above. 

\smallskip
{\csc Remark 2.} 
A sometimes unfortunate side effect of additive expansion 
procedures like the one developed above is that the 
density estimator occasionally may become negative. 
Thus $\hatt S_0(\hatt\sigma^{-1}h)$ and $\hatt T_0(\hatt\sigma^{-1}h)$
above are formed employing integrands involving 
say $\hatt g_{2m}(y)$ that once in a while is negative. 
While this is annoying from a density estimation point of view,
like the situation with higher order kernels,  
it does not disturb our method. If $\hatt g_{2m}$ is negative
in places a remedy is to lift it up the entire curve the required amount.
But this fortunately does not change the minimiser of (4.5) at all. 
Another view supporting the notion that occasional negativity 
of the density estimator is not a disturbing matter 
for the present method is as follows. 
As discussed in the following section, 
the precision of a method for finding the best $h$ 
is essentially determined by the implicit estimate of the 
fourth $g$-derivative at zero. Thus the actual level of 
the estimated density is unimportant, and, specifically,
it is immaterial whether the density has been lifted or not. 
\eject 

\smallskip
{\csc Remark 3.}
There are alternative estimators for $\alpha_{2j}$ 
that are sometimes favourable, having to do with 
separate treatment of the diagonal terms.
Some initial discussion is given in Section 6. 

\section 
\centerline{\bf 5. Analysing behaviour of the new h selectors}

\hop 
This section analyses properties of 
semiparametric bandwidth selection methods 
that employ the Hermite machinery of the previous section. 
The results will also be used in Section 6 to 
suggest various amendments and modifications to the direct method 
that uses minimisation of (4.5). 

\subsection
{\sl 5.1. Background on competing methods.} 
Estimating $h_n$ of (1.2) must for large $n$ 
in one way or another be related to estimating 
the quantity $R(f'')$, see e.g.~(1.3) and (3.4) above. 
Some of the best estimators in the literature for this 
crucial parameter use kernel smoothing in various forms.
The nature of our own results, to be described below, 
makes it relevant to describe some features of these other constructions. 
Hall and Marron (1987) gave estimators with 
bias $O(\beta^{2r})$ and 
$${\rm variance}={4\over n}\Bigl\{\int (f^{(4)})^2f\,\d x-R(f'')^2\Bigr\}
        +O(\beta^2n^{-1}+(n^2\beta^9)^{-1}), \eqno(5.1)$$
in terms of a bandwidth parameter $\beta$
for a kernel $L$ of order $2r$, that is, 
$k_j(L)=\int u^jL(u)\,\d u$ is zero for $j=1,\ldots,2r-1$ 
while $k_{2r}(L)>0$. See also Bickel and Ritov (1988). 
It turns out that the bias term here is always negative,
so instead of choosing $\beta$ to balance squared bias with variance, 
Jones and Sheather (1991) `added back bias' 
using suitable extra `diagonal terms' in a modified estimator construction. 
Their estimator, say $\hatt R_D(\beta)$, 
using an ordinary second order kernel, has 
$$\E\hatt R_D(\beta)=R(f'')-\half\beta^2k_2(L)R(f''')+L^{(4)}(0)(n\beta)^{-5}
        +o(h^2) $$
and variance of the same size as given above for 
the Hall and Marron estimators. 
This was successfully utilised in Sheather and Jones (1991) 
to construct an $h$ selector which appears to be one of
the very best in existence, cf.~extensive simulations 
in Jones, Marron and Sheather (1996). 
It works as follows. The leading terms of the bias 
cancel out when $\beta$ is a certain $c_0n^{-1/7}$, say, 
which they re-express as $ch_{n,a}^{5/7}$ using the \amise-optimal 
$h_{n,a}$ given in (3.4). 
This invites putting $\hatt\beta(h)=\hatt ch^{5/7}$, say, 
in the equation governing the final choice of $h$.
Here $\hatt c$ involves first stage estimation of both 
$R(f'')$ and $R(f''')$ but with only lower-level precision required,
see their paper for details. The final equation determining 
their recommended $\hatt h_{\rm SJ}$ is 
$$h=\{R(K)/k_2^2\}^{1/5}\hatt R_D(\hatt ch^{5/7})^{-1/5}
        n^{-1/5}. \eqno(5.2) $$

\subsection
{\sl 5.2. Bias for the Hermite approximation rules.}
As a benchmark we first consider the normal reference distribution,   
for which we find 
$$g^{(4)}(0)=R(N(\mu,\sigma^2)'')=3\sigma^{-5}/(8\sqrt{\pi}); \eqno(5.3)$$ 
this is what leads to the ordinary form of the normal reference rule.
We will see how the Hermite expansion model (4.2) 
is able to make its way from the normal-based approximation (5.3)
to the completely general $R(f'')$, 
through a combination of choosing $m$ larger and $\hhh$ smaller. 
In other words, the approximation that at each stage
involves only a finite and typically small number of parameters 
is able to bridge the full gap 
from simplistic parametrics to full nonparametrics. 

The Hermite model density can be written 
$(\sqrt{2}\sigma)^{-1}g_{0,2m}((\sqrt{2}\sigma)^{-1}y)$, 
where $g_{0,2m}(y)$ is 
$\phi(y)\sum_{j=0}^m\{\alpha_{2j}/(2j)!\}\,H_{2j}(\hhh^{-1}y)$. 
Hence its fourth derivative at zero becomes 
$(\sqrt{2}\sigma)^{-5}\allowbreak g_{0,2m}^{(4)}(0)$. 
And the fourth derivative of $g_{0,2m}$ at zero can be written
$$\sum_{j=0}^m{\alpha_{2j}\over (2j)!}
        \{\phi^{(4)}(0)H_{2j}(0)+6\phi''(0)H_{2j}''(0)/\hhh^2
        +\phi(0)H_{2j}^{(4)}(0)/\hhh^4\}. $$
This can be nicely simplified using various 
detectable facts about Hermite polynomials. 
The first fact is that $H_{2j}''=2j(2j-1)H_{2j-2}''$. 
Furthermore, the successive values of $H_{2j}(0)$ are
$1,-1,3,-3\cdot5,3\cdot5\cdot7,-3\cdot5\cdot7\cdot9$ and so on. 
This leads after some further simplifications to 
an expression for the central quantity of the form
$$\eqalign{g_{2m}^{(4)}(0)
&={3\sigma^{-5}\over 8\sqrt{\pi}}
        \Bigl[\alpha_0+
        \sum_{j=1}^m{\alpha_{2j}\over 2^j j!}(-1)^j
        \{1+4j/\hhh^2 + (4/3)j(j-1)/\hhh^4\}\Bigr] \cr
&={3\over (\sqrt{2}\sigma)^5}\sum_{j=0}^m{\alpha_{2j}\over (2\pi)^{1/2}}
        {(-1)^j\over 2^j j!}\{1+4j/\hhh^2+(4/3)j(j-1)/\hhh^4\}. \cr}
                        \eqno(5.4)$$
This also illustrates one way in which 
the normal reference rule can be corrected for non-zero coefficients. 
Note that this method cleverly estimates $R(f'')$ 
without having to use odd-numbered coefficients. 

We are now in position to reach the following result.
The proof is relegated to the Appendix. 

{\smallskip\sl
{\csc Bias Proposition.} 
For the expansion model (4.2) of order $2m\ge4$ 
and Hermite bandwidth $\hhh$, let $R_{2m}=g_{2m}^{(4)}(0)$ 
be the implicit approximation of $R=R(f'')$. 
Assume that $g$ has finite variance and $2m+2$ derivatives at zero. 
Then, as $\hhh$ gets small, the following holds:
$$R_{2m}=R(f'')+(-1)^m{1\over (\sqrt{2}\sigma)^5}
  {G^{(2m+2)}(0)\over 2^{m-1}(m-1)!}\,\hhh^{2m-2}+o(\hhh^{2m-2}), $$
where $G(x)$ is the function 
$\exp(\half x^2)g(\sqrt{2}\sigma x)\sqrt{2}\sigma$. 
\smallskip}
\eject

The error term here is $O(\hhh^{2m})$ if $g$ has an additional two
derivatives at zero. 
Note the similarity of this result to those available 
for the $R(f'')$ estimators mentioned in Section 5.1. 

To find out more about the size of the modelling bias involved,
let us write $g(y)=(\sqrt{2}\sigma)^{-1}g_0((\sqrt{2}\sigma)^{-1}y)$
in terms of $g_0$, the normalised $Y$ density with variance one. 
We need successive derivatives of the $G(x)$ function at zero,
and to this end note that the $(2i)$th derivative of zero
of the $\exp(\half x^2)$ function can be recognised from 
Hermite polynomial coefficients, and is $(2i)!/(2^i i!)$. 
The odd ones disappear. Hence 
$$G^{(2j)}(0)=\sum_{i=0}^j{2j\choose 2i}{(2j-2i)!\over 2^{j-i}(j-i)!}
        g^{(2i)}(0)(\sqrt{2}\sigma)^{2i+1}
        =\sum_{i=0}^j{(2j)!\over (2i)!2^{j-i}(j-i)!}
        g_0^{(2i)}(0). $$
The coefficients here are precisely of the form $w(2j,j-i)$,  
where these are defined and used in the proof of the bias proposition,
see the Appendix, and are those entering Hermite polynomials, 
but here without sign corrections. 
Thus the implicit modelling bias involved when using the (4.2) 
expansion is seen via 
$$R_{2m}\doteq R(f'')+{1\over (\sqrt{2}\sigma)^5}
        \Bigl\{{(-1)^m\over 2^{m-1} (m-1)!}
        \sum_{i=0}^{m+1}w(2m+2,2m+2-2i)g_0^{(2i)}(0)\Bigr\}
        \,\hhh^{2m-2}, \eqno(5.5)$$
the error term being $o(\hhh^{2m-2})$ if $g$ has $2m-2$ derivatives
at zero and actually $O(\hhh^{2m})$ if it possesses an additional two. 
For example, the approximate model biases 
when using respectively coefficients up to order four, 
up to order six and up to order eight are 
$$\eqalign{
{\rm bias}_4&\doteq {1\over 2}{G^{(6)}(0)\over (\sqrt{2}\sigma)^5}\,\hhh^2
        ={1\over 2}{1\over (\sqrt{2}\sigma)^5}
        ( 15g_0
          +45g_0^{(2)} 
          +15g_0^{(4)}
          +g_0^{(6)})\,\hhh^2, \cr
{\rm bias}_6&\doteq-{1\over 8}{G^{(8)}(0)\over (\sqrt{2}\sigma)^5}\,\hhh^4 \cr 
&=-{1\over 8}{1\over (\sqrt{2}\sigma)^5}
        (105g_0
         +420g_0^{(2)}
         +210g_0^{(4)}
         + 28g_0^{(6)}
          +  g_0^{(8)})\,\hhh^4, \cr 
{\rm bias}_8&\doteq{1\over 48}{G^{(10)}(0)\over (\sqrt{2}\sigma)^5}\,\hhh^6 \cr
&={1\over 48}{1\over (\sqrt{2}\sigma)^5}
        (945g_0
        +4725g_0^{(2)}
        +3150g_0^{(4)}
         +630g_0^{(6)}
          +45g_0^{(8)}
           + g_0^{(10)})\,\hhh^6 \cr}$$
(the functions being evaluated at zero), 
smoothness of $g_0$ around zero permitting. 
Thus estimating only four coefficients from the data
already has the potential of the small bias of size $\hhh^6$,
estimation of five coefficients gives the potentially 
very small bias of size $\hhh^8$, and so on.
It is helpful to note that all these complicated-looking 
bias term expressions are actually zeroed under normality, 
also indicating that they will not be too big if $g$ is within 
a reasonable vicinity of the normal. See also Section 6.

\subsection 
{\sl 5.3. Relative variability of the Hermite rules.} 
Implicit in our Hermite $h$ selection scheme is an 
estimator of the fourth $g$-derivative at zero, 
which by (5.4) and (4.4) can be expressed as 
$$\eqalign{\hatt R_{2m}
&={3\over (\sqrt{2}\sigma)^5}
        \sum_{j=0}^m {\hatt\alpha_{2j}\over (2\pi)^{1/2}}
        {(-1)^j\over 2^j j!}\{1+4j/\hhh^2+(4/3)j(j-1)/\hhh^4\} \cr
&={3\over (\sqrt{2}\sigma)^5}
{n\choose 2}^{-1}\sum_{i<l}\Bigl\{\sum_{j=0}^m 
        \hhh^{-1}p_{2j}\Bigl({\hhh^{-1}Y_{i,l}\over \sqrt{2}\sigma}\Bigr)
                c(j)\Bigr\}, \cr} \eqno(5.6)$$
in which $c(j)=(-1)^j\{1+4j/\hhh^2+(4/3)j(j-1)/\hhh^4\}/(2^j j!)$
and $p_{2j}(u)$ is the function $H_{2j}(u)\phi(u)\exp(\half \hhh^2 u^2)$.
As noted after (4.4) it is not a necessity to use 
$\hatt\sigma$ estimated from data, the $\sigma$ can be a pre-determined value. 
And it is easier to work with a fixed $\sigma$ when 
attacking the variance question. 
The quite lengthy proof of the following result 
has been sent to the Appendix. 

{\smallskip\sl
{\csc Variance Proposition.} 
The variance of $\hatt R_{2m}$ can be expressed as 
$${4\over n}\Bigl\{\int (f^{(4)})^2f\,\d x-R(f'')^2\Bigr\}
        +O(n^{-1}\hhh^2+(n^2\hhh^9)^{-1}) $$
when $\hhh$ goes to zero slowly enough 
to let $n\hhh^9$ grow to infinity. 
\smallskip}

Again this is quite and perhaps even surprisingly similar 
to results for competing estimators mentioned in Section 5.1. 
Importantly, the $n^{-1}$ constant here is also 
the theoretically best possible, in a precise asymptotic sense; 
this may be shown following results of Fan and Marron (1992). 

\section 
\centerline{\bf 6. Combating the bias}

\hop  
Some simulation experience with the rules associated with
minimisation of (4.5), working with finite normal mixtures 
of the Marron and Wand (1992) variety, 
suggested that the selected $\hatt h$ values 
tended to have low variability but that they too often erred 
on the positive side of the target $h_n$. 
Examination revealed that this was caused 
by underestimation of $R(f'')$ when using $\hatt R_{2m}$ 
in its direct form (5.6). In other words, 
in the notation of Section 5.2,
the $G^{(6)}(0)$ quantity was typically negative,
while $G^{(8)}(0)$ was positive, and so on. 

One may consider several possible routes towards combating  
this negative bias. Let us write 
$$\E\hatt R_{2m}=R(f'')+(-1)^m\{(\sqrt{2}\sigma)^5 2^{m-1}m!\}^{-1}
        b_{2m}\hhh^{2m-2}+o(\hhh^{2m}), \eqno(6.1)$$
using the bias proposition, with $b_{2m}=G^{(2m+2)}(0)$. 
Results of Section 5 entail that $b_{2m}$ may be expressed in
terms of $g$'s even derivatives at zero, in other words  
in terms of $R(f),R(f'),\ldots,R(f^{(m+1)})$, 
and an estimator can be constructed for $b_{2m}$ based on this. 
It is simpler however to use the expansion methods already
worked with above. Some analysis shows that 
$$G^{(2m+2)}(0)=(2\pi)^{-1/2}(\alpha_{2m+2}-\half\alpha_{2m+4}
        +\hbox{$1\over8$}\alpha_{2m+6}-\cdots)/\tilda\hhh^{2m+2}, \eqno(6.2)$$
in terms of a long enough expansion with a new and typically 
bigger Hermite bandwidth $\tilda\hhh$ and associated 
$\alpha_{2j}$ coefficients. Thus a simple estimate of 
$b_{2m}$ is $(2\pi)^{-1/2}\tilda\alpha_{2m+2}/\tilda\hhh^{2m+2}$, 
where $\tilda\alpha_{2m+2}$ is as in (4.4) but with new 
bandwidth $\tilda\hhh$. 

Another proposal, motivated by the success of the Jones and Sheather 
method, is to use a `diagonals-in' sister version of estimator (5.6). 
That is, 
$$\eqalign{\hatt R_{2m,D}
&={3\over (\sqrt{2}\sigma)^5}
{1\over n^2}\sum_{i=1}^n\sum_{l=1}^n \Bigl\{\sum_{j=0}^m 
        \hhh^{-1}p_{2j}\Bigl({\hhh^{-1}Y_{i,l}\over \sqrt{2}\sigma}\Bigr)
                c(j)\Bigr\} \cr 
&=(1-n^{-1})\hatt R_{2m} \cr 
&\qquad +{1\over n\hhh^5}
        {3\over (\sqrt{2}\sigma)^5}{1\over (2\pi)^{1/2}}  
        \sum_{j=0}^m{(2j)!\over 2^{2j} (j!)^2}
        \{(4/3)j(j-1)+4j\hhh^2+\hhh^4\}. \cr}$$
Thus the added quantity is positive, which is good in view 
of the non-negligible negative bias that bothers $\hatt R_{2m}$. 
The idea is to choose $\hhh$ to make the leading terms
of the bias approximately cancel each other. 

It is clear that there must be a great variety of co-existing 
natural specialisations of these ideas, and it is not at all 
clear a priori which will perform better than others 
in various situations. Here we are content to describe 
only some representatives that look promising on the
basis of parallel experience for kernel based methods. 

\smallskip
{\csc Proposal 1:} 
In view of the Taylor approximation to $\hatt q(h)$
given in Section 3.5, let $\hatt h_{\rm I}$ minimise 
$$\eqalign{(nh)^{-1}R(K)
&+\quart k_2^2h^4\{\hatt R_4(\hhh)
        -(2\hatt\tau^5)^{-1}\hatt b_4\hhh^2\}(1-n^{-1}) \cr
&\qquad\qquad 
        +\twentyfourth k_2k_4h^6\hatt S_6(\tilda\hhh)(1-n^{-1}), \cr}
                        \eqno(6.3)$$
where $\hatt S_6(\tilda\hhh)$ is an estimate of $g^{(6)}(0)=-R(f''')$
formed as for $R(f'')$, but using terms of order 0, 2, 4, 6,
and using another and bigger pilot bandwidth $\tilda\hhh$. 
Here $\tau$ and $\hatt\tau$ are used for $\sqrt{2}\sigma$ 
and $\sqrt{2}\hatt\sigma$, for simplicity, and 
$\hatt b_4=(2\pi)^{-1/2}\tilda\alpha_6/\tilda\hhh^6$.
A suitable pilot value for $\tilda\hhh$ needs to be decided on. 
One might also try to deduct bias for $\hatt S_6(\tilda\hhh)$ too,
although this seems less crucial.
One might drop the $h^6$ term here, aiming more directly 
at the approximate $h_{n,a}$ minimiser, but keeping it in
is more in the spirit of our proclaimed non-asymptotic view 
using $q(h)$. One may also consider the option of 
putting $\hhh=\hatt ch^{5/7}$ here, 
with $\hatt c$ as given in the course of Proposal 2, 
and then minimise the resulting expression over $h$.
This would avoid having to select a pilot parameter value for $\hhh$. 

\smallskip
{\csc Proposal 2:} 
We now present an attempt to follow the chain of arguments 
and calculations used so successfully by Sheather and Jones,
but exploiting our identity of Section 2 rather than 
large-sample approximations, and using Hermite expansions
rather than kernel estimators as auxiliary machinery. 
Consider in general terms 
$$\hatt g_{2m,D}(y,\hhh)=\phi\Bigl({y\over \tau}\Bigr){1\over \tau}
        \sum_{j=0}^m{\hatt\alpha_{2j,D}\over (2j)!}
        H_{2j}\Bigl({y\over \hhh\tau}\Bigr), $$
estimated from the (4.2) model, but now using `diagonals in' 
versions of coefficient estimates. Then its fourth derivative 
at zero is precisely $\hatt R_{2m,D}$ considered above. 
A formula for $\hatt q_D(h,\hhh)=\int A_K(v)\hatt g_{2m,D}(hv,\hhh)\,\d v$
is already available in Section 4, and indeed it makes sense 
to minimise the accompanying version of equation (4.5),
for a suitable pilot value of $\hhh$. It is more in spirit
of Sheather and Jones' final recommendation, however, to 
use a special $\hhh$ that depends on $h$, as follows. 
Its motivation is that the leading terms of the bias of $\hatt R_{4,D}$ are
$$\half {1\over \tau^5}b_4\hhh^2+{1\over n\hhh^5}{3\over \tau^5}
        {1\over (2\pi)^{1/2}}(1+5\hhh^2+(15/8)\hhh^4), $$
and these are made to cancel if $\hhh=J_n(b_4)n^{-1/7}$, say,
defined as the solution to 
$$\hhh=\{6/(2\pi)^{1/2}\}^{1/7}(-1/b_4)^{1/7}
        (1+5\hhh^2+(15/8)\hhh^4)^{1/7}n^{-1/7}. $$
Here one takes care to use a negative estimate of $b_4=G^{(6)}(0)$. 
A quick approximation is $\hhh=\{6/(2\pi)^{1/2}\}^{1/7}(-1/b_4)^{1/7}n^{-1/7}$.
This can be written in the form $\hhh=c_nh_{n,a}^{5/7}$.
Estimating $b_4$ as in Proposal 1 gives a suitable $\hatt c_n$.
Now define $\hatt h_{\rm II}$ to be the minimiser of 
$$\eqalign{
\dna_{\rm II}(h)&=(nh)^{-1}R(K)+\hatt q_D(h,\hatt\hhh(h)) \cr
&=(nh)^{-1}R(K)+\int A_K(v)\hatt g_{2m,D}(hv,\hatt c h^{5/7})\,\d v. \cr}$$
For a normal kernel this is exactly formula (4.5),
but with $\hatt c h^{5/7}$ replacing $\hhh$ there, 
and with diagonals-in versions 
$$\hatt\alpha_{2j,D}=(1-n^{-1})\hatt\alpha_{2j}
        +(n\hhh)^{-1}(-1)^j (2j)!/(2^j j!). $$
An easy approximation, which might suffice in practice, is  
$$\hatt c=\Bigl\{{6/(2\pi)^{1/2})\over R(K)/k_2^2}\Bigr\}^{1/7}
        {\hatt R(f'')^{1/7}\over (-\hatt b_4)^{1/7}}. $$
This proposal needs for its completion a sound and robust
pilot estimate of $b_4$. 

\smallskip
{\csc Proposal 3:} 
The methods above used very short expansions 
containing $\alpha$ coefficients up to order $2m=4$. 
We now try using $\alpha$s up to order $2m=6$. 
Then the parallel of Proposal 1 is to minimise 
$$\eqalign{(nh)^{-1}R(K)
&+\quart k_2^2h^4\{\hatt R_6(\hhh)
        -(8\hatt\tau^5)^{-1}\hatt b_6\hhh^4\}(1-n^{-1}) \cr 
&\qquad\qquad 
        +\twentyfourth k_2k_4h^6\hatt S_8(\tilda\hhh)(1-n^{-1}), \cr}$$
where $\hatt b_6=(2\pi)^{-1/2}\tilda\alpha_8/\tilda\hhh^8$.
There would be a parallel to Proposal 2, with some extra work. 
\eject 

\smallskip
{\csc Proposal 4:}
Yet another idea is to plug in the best Hermite estimates one can 
think of for $R(f'')$ and $R(f''')$, in the approximation formula for 
$h_n$ given in Fan and Marron (1992). 

\smallskip 
{\csc Remark.} 
It is important to study the large-sample behaviour 
of the explicit and implicit $h$ selectors here, 
to understand better how well they perform, and to compare 
with other competing methods. This will be tended to in
future work. 

\section 
\centerline{\bf 7. Concluding comments} 

\hop 
\subsection
{\sl 7.1. Finite-sample modification of the Sheather--Jones rule?}
The Sheather and Jones (1991) rule in effect uses 
$\hatt S(\beta)=\hatt g_D^{(4)}(0,\beta)$ to estimate $R(f'')$,
where $\hatt g_D(y,\beta)$ is a diagonals-in density kernel density 
estimate with bandwidth $\beta$. 
A `finite-sample corrected' version of this, 
in the spirit of other methods developed here, 
would be as follows. Let $\hatt\beta(h)=\hatt c h^{5/7}$
where $\hatt c$ is constructed in suitable parallel 
to a similar object in Sheather and Jones (1991).
Then let $\hatt h$ minimise 
$$\dna_{\rm SJ}(h)=(nh)^{-1}R(K)
        +\int A_K(v)\hatt g(hv,\hatt\beta(h))\,\d v. $$
This might be worth pursuing.       

\subsection
{\sl 7.2. Giving the $g$ estimator a normal start.} 
Whereas the \ucv{} method uses the empirical distribution
of $Y_{i,j}$s to estimate $q(h)=\E\,h^{-1}A_K(h^{-1}Y)$,
the \scv{} method, viewed in the light of this paper, 
uses a smoother density estimate,
giving $\hatt q(h)=\int h^{-1}A_K(h^{-1}y)\hatt g(y)\,\d y$.
There are other versions of this that could easily perform
better, through exploitation of the prior knowledge 
that $g(y)$ is often not far from being normal. 
One density estimator which can take such knowledge into account
is the multiplicative method of Hjort and Glad (1995).
Here it takes the form
$$\hatt g(y)=\phi\Bigl({y\over \sqrt{2}\hatt\sigma}\Bigr)
        {1\over \sqrt{2}\hatt\sigma}
        {1\over n(n-1)}\sum_{i\not=j} L_{\tilda h}(Y_{i,j}-y)\Big/
        \phi\Bigl({Y_{i,j}\over \sqrt{2}\hatt\sigma}\Bigr)
                {1\over \sqrt{2}\hatt\sigma} $$
in terms of a bandwidth parameter $\tilda h$. 
The intended $h$ selection mechanism is to let $\hatt h$ minimise 
$$\hatt{\dna}(h)=(nh)^{-1}R(K)+\hatt q(h)
        =(nh)^{-1}R(K)+(1-n^{-1})\hatt S(h)-2\hatt T(h), $$
where $\hatt S(h)=\int g_K(v)\hatt g(hv)\,\d v$ and 
$\hatt T(h)=\int K(v)\hatt g(hv)\,\d v$. With a normal kernel for $L$ 
this can be seen to yield 
$$\eqalign{
\hatt S(h)&={n\choose 2}^{-1}\sum_{i<j}
        {1\over (2\pi)^{1/2}}{\tilda\sigma_2\over \sqrt{2}h}
        \exp\Bigl\{-\half\Bigl({Y_{i,j}\over \tilda h}\Bigr)^2
        \Bigl(1-{h^2\tilda\sigma_2^2\over \tilda h^2}
        -{\tilda h^2\over 2\sigma^2}\Bigr)\Bigr\}, \cr
\hatt T(h)&={n\choose 2}^{-1}\sum_{i<j}
        {1\over (2\pi)^{1/2}}{\tilda\sigma_1\over h}
        \exp\Bigl\{-\half\Bigl({Y_{i,j}\over \tilda h}\Bigr)^2
        \Bigl(1-{h^2\tilda\sigma_1^2\over \tilda h^2}
        -{\tilda h^2\over 2\sigma^2}\Bigr)\Bigr\}, \cr}$$
in which 
$$1/\tilda\sigma_1^2=1+h^2/(2\sigma^2)+h^2/\tilda h^2
        \quad {\rm and} \quad 
  1/\tilda\sigma_2^2=1/2+h^2/(2\sigma^2)+h^2/\tilda h^2. $$

Note that $\tilda h\arr0$ gives \ucv, 
that $\sigma\arr\infty$ gives \scv,
and that even $\ucv$ emerges if only $\tilda h\arr0$ 
for fixed $\sigma$.

An alternative estimator of $R(f'')=g^{(4)}(0)$ emerges, 
as a separate bonus of this approach. It takes the form
$$\hatt R={n\choose 2}^{-1}\sum_{i<j}
   \exp\Bigl(\half {Y_{i,j}^2\over \tau^2}\Bigr)
\Bigl\{
 {3\over \tau^4}L\Bigl({Y_{i,j}\over \tilda h}\Bigr){1\over \tilda h}
-{6\over \tau^2}L''\Bigl({Y_{i,j}\over \tilda h}\Bigr){1\over \tilda h^3}
+{1\over \tau}L^{(4)}\Bigl({Y_{i,j}\over \tilda h}\Bigr){1\over \tilda h^5}
\Bigr\}, $$
where $\tau=\sqrt{2}\hatt\sigma$. Some additional work shows 
that also this estimator has variance of the optimal form (5.1), 
while its bias can be written 
$$\E\hatt R-R(f'')=\half k_2(L)\tilda h^2(g_{\rm in}r^{(2)})^{(4)}(0)
        +\twentyfourth k_4(L)\tilda h^4(g_{\rm in}r^{(4)})^{(4)}(0)+o(h^4), $$
where $g_{\rm in}$ is the initial best normal approximation,
that is, the $N(0,2\sigma^2)$ density, and $r=g/g_{\rm in}$. 
The bias here is potentially smaller than that of 
the more traditional kernel estimator that 
`starts with nothing' rather than starting with a normal approximation. 
In particular the bias is close to zero when $f$ is normal. 
Separate $h$ selectors can be constructed using $\hatt R$. 

This could be implemented and tried out. 
A good pilot value for $\tilde h$ could be worked out
from calculations that to some extent would parallel 
those of Hjort and Glad (1995). 

\subsection
{\sl 7.3. A local likelihood approach.}
The Hermite methods are essentially additive.
Here we take a look at methods that are multiplicative in nature. 
As a local parametric model for $g(t)$, with $t$ in the 
vicinity of a fixed $y$, employ a suitable $g(t,\theta)$. 
The local parameters $\theta$ are estimated by maximising 
$${1\over n(n-1)}\sum_{i\not=j} L_b(y_{i,j}-y)\log g(y_{i,j},\theta)
        -\int L_b(t-y)g(t,\theta)\,\d t, $$
where $b$ is a bandwidth parameter for the kernel $L$. 
This is the local likelihood approach of Hjort and Jones (1996)
and Loader (1996), here using a broad enough model 
for its fourth derivative at zero to be meaningfully expressed. 
The $Y_{i,j}$ data are partly dependent, but that does not 
disturb the basic motivation behind the method,
which still aims at the best local parametric approximation
in a suitable local Kullback--Leibler distance fashion. 
Here we aim directly at zero, around which $g$ is symmetric,
so a natural local model could be 
$\exp(a+bt^2+ct^4)$. The procedure would then 
take the following form: maximise 
$${n\choose2}^{-1}\sum_{i<j}L_h(y_{i,j})
        (a+by_{i,j}^2+cy_{i,j}^4)
        -\int L_h(t)\exp(a+bt^2+ct^4)\,\d t $$
with respect to the three parameters. This gives the estimate 
$$R^*=g^{(4)}(0,\hatt a,\hatt b,\hatt c)
        =\exp(\hatt a)(24\hatt c + 12\hatt b^2). $$
The kernel $L$ should have bounded support here. 

\subsection
{\sl 7.4. Density estimates of g.}
The difference density $g$ is instrumental in several 
of our ingredients. Estimating $g$ also has separate interest,
and the following might come in handy on a rainy day. Consider 
$$\hatt g(y)={1\over n(n-1)}\sum_{i\not=j}L_h(Y_{i,j}-y)
        ={n\choose2}^{-1}\sum_{i<j}\bar L_h(Y_{i,j},y), $$
in which $\bar L_h(y_{i,j},y)=\half\{L_h(y_{i,j}-y)+L_h(y_{i,j}+y)\}$. 
This is one of two canonical kernel density estimators for $g$.
The other one comes from $g(y)=\int f(x)f(x+y)\,\d x$ and uses 
$$\eqalign{\hatt g_D(y)
&=\int\hatt f(x)\hatt f(x+y)\,\d x \cr 
&={1\over n^2}\sum_{i,j}\int K_h(x-x_i)K_h(x+y-x_j)\,\d x \cr 
&={1\over n^2}\sum_{i,j}(K_h*K_h)(y_{i,j}-y), \cr} $$
leaving diagonal contributions in. 

Let us analyse the first one. Its mean is 
$$\eqalign{
e_h(y)&=\E\bar L_h(Y_{i,j},y) \cr 
&=\half\int\{L_h(z-y)+L_h(z+y)\}g(z)\,\d z \cr 
&=\half\int L(u)\{g(y+hu)+g(y-hu)\}\,\d u, \cr}$$
which becomes of the familiar $g+\half\lambda_2h^2g''+\cdots$ kind.
The variance is more difficult. 
From well-known results on $U$-statistics,  
see e.g.~Serfling (1980, Ch.~5), we have 
$$\Var\,\hatt g(y)={4\over n}{n-2\over n-1}{\rm cov}_0(y)
        +{2\over n(n-1)}\Var_0(y), $$
where $\Var_0(y)$ is the variance of $\bar L_h(Y_{1,2},y)$
and ${\rm cov}_0(y)$ is the covariance between 
$\bar L_h(Y_{1,2},\allowbreak y)$ and $\bar L_h(Y_{1,3},y)$.
The variance term can be written
$$\eqalign{
\quart\int&\{h^{-1}L(h^{-1}(z-y))+h^{-1}L(h^{-1}(z+y))\}^2g(z)\,\d z
        -e_h(y)^2 \cr 
&=\quart h^{-1}\int L(u)^2\{g(y+hu)+g(y-hu)\}\,\d u \cr 
&\qquad\qquad 
+\half h^{-1}\int L(u)L(u+2y/h)g(y+hu)\,\d u -e_h(y)^2. \cr}$$
The covariance term becomes 
$$\eqalign{
&\quart h^{-2}{\rm cov}\{L(h^{-1}(Y_{1,2}-y))+L(h^{-1}(Y_{1,2}+y)), \cr 
&\qquad\qquad
        L(h^{-1}(Y_{1,3}-y))+L(h^{-1}(Y_{1,3}+y))\} \cr
&=\quart h^{-2}\int\int\{L(h^{-1}(z_1-y))+L(h^{-1}(z_2-y))
        +L(h^{-1}(z_1-y))L(h^{-1}(z_2+y)) \cr 
&       +L(h^{-1}(z_1+y))L(h^{-1}(z_2-y))
        +L(h^{-1}(z_1+y))L(h^{-1}(z_2+y))\} \cr 
&\qquad\qquad 
        \bar g(z_1,z_2)\,\d z_1\,\d z_2 - e_h(y)^2 \cr
&=\quart\int\int L(u_1)L(u_2)
        \{\bar g(y+hu_1,y+hu_2) + \bar g(y+hu_1,-y+hu_2) \cr 
&\qquad\qquad   
+\bar g(-y+hu_1,y+hu_2) + \bar g(-y+hu_1,-y+hu_2)\}\,\d u_1\,\d u_2
        -e_h(y)^2, \cr}$$
in which $\bar g(z_1,z_2)$ is the simultaneous density 
for two related differences $(X_2-X_1,X_3-X_1)$. 
It follows that 
$$\eqalign{\Var\,\hatt g(y)
&={4\over n}{n-2\over n-1}g^*(y)
        +{R(L)\over n(n-1)h}g(y) \cr 
&\qquad\qquad
        -{4n-6\over n(n-1)}e_h(y)^2+O(h^2/n)+O(n^{-2}h^{-1}g_L(2y/h)), \cr}$$
where $g^*(y)$ is the symmetrised 
$(1/4)\{\bar g(y,y)+\bar g(y,-y)+\bar g(-y,y)+\bar g(-y,-y)\}$.

There are a couple of points worth discussing briefly here.
The first is that while the traditional $U$-statistics result
says that the covariance term is most important
and will dominate the variance term for large $n$,
this does not happen here, due to the presence of the $h$
which approaches zero with growing $n$. 
The covariance terms contribution is $O(n^{-1})$,
as is that of the subtracted $e_h^2$ term,
and the variance terms contribute markedly,
namely $O((n^2h)^{-1})$, but this is still dominated by 
the $O(n^{-1})$ term if only $nh\arr\infty$. 
Accordingly, $g$ can be estimated with $1/n$ precision,
unlike $f$, which can only be estimated with $1/n^{4/5}$ precision. 

\subsection
{\sl 7.5. Correcting for longer tails.} 
Suppose $Y$ has somewhat longer tails than predicted
by the normal, thus making the normal reference rule less
than perfect. One way of correcting for such behaviour of 
data is via Hermite expansions again, with longtailedness
showing up suitably in the coefficients,  
Another possibility is to model the density as a $t$ density: 
$${Y\over \sqrt{2}\sigma}\sim \Bigl({\nu-2\over \nu}\Bigr)^{1/2}t_\nu
        \sim (\nu-2)^{1/2}N(0,1)/(\chi^2_\nu)^{1/2}. $$
We write it in this form since the variance of $Y/(\sqrt{2}\sigma)$
must be one. The procedure is to estimate $\nu$ from data,
thus leading to a suitable $\hatt q(h)$ by insertion in formula (2.4).
The bandwidth to use in the end is the one minimising 
the consequent (3.1). Numerical integration seems necessary,
but is not a serious obstacle in practice. 

There are a couple of ways of estimating $\nu$ from the $Y_{i,j}=X_j-X_i$
data. One is to set the average $\hatt\lambda_4$ 
of all $Y_{i,j}^4/(4\hatt\sigma^4)$
equal to the value predicted by the $t$ model, that is,
to $3\{1+2/(\nu-4)\}$. If the observed value of $\hatt\lambda_4$ 
is smaller than 3, then go back to normal reference after all. 
A second possibility is based on first finding $z_0$, 
the median of all $|Y_{i,j}|/(\sqrt{2}\hatt\sigma)$.
Thus 50\% of the standardised differences are found in $(-z_0,z_0)$. 
Then determine $\nu$ from $G_\nu((\nu/(\nu-2))^{1/2}z_0)={3\over4}$,
where $G_\nu$ is the cumulative distribution function for 
a $t_\nu$. 

\subsection
{\sl 7.6. Omitting non-significant coefficients.}
Another idea to explore is to test each hypothesis 
$H_0\colon\alpha_{2j}=0$, at suitable significance levels, 
and only include those with a clearly visible presence. 

\subsection
{\sl 7.7. Rules using Hermite expansion for $f$.} 
Our Hermite-based rules are based on expansions for 
the difference density $g$, exploiting the fact that 
other aspects of the density $f$ simply do not matter 
for the \mise, as seen from the proposition of Section 2. 
One could also work out expansions for $f$ instead of $g$,
still along the lines of Section 4. Some such was in fact 
carried out in Hjort and Jones (1995); see also Exercises
9, 20, 21 in the collection of Hjort (1993). 
The approach of the present paper 
is more immediately appealing and elegant, however,
in that it actively exploits and benefits from the symmetry of $g$. 

\subsection
{\sl 7.8. A theoretical question to ponder.} 
We saw that $g$ is more precisely estimated than $f$. 
Presumably also the density $p_2$ of $(X_1+X_2)/\sqrt{2}$
can be estimated with $1/n$ precision, using a kernel 
estimate for all observed sums of pairs. And, presumably,
the density $p_k$ of $(X_1+\cdots+X_k)/\sqrt{k}$ becomes
more and more smooth and can be estimated better and better,
as $k=3,4,5,\ldots$ grows. It would have theoretical interest
to pinpoint better this bridge into smoothness and the 
domain of the central limit theorem. 

\subsection
{\sl 7.9. A World Cup contest.} 
It would also be interesting to pit the many different 
estimators of $R_2=\int(f'')^2\,\d x$ against each other, 
for a variety of estimands $f$. The $R_2$ quantity 
is crucial for the smoothness problem, as we have seen,
and also has independent interpretation as `density roughness'. 

\section 
\centerline{\bf Appendix: Bias and variance for the Hermite method}

\hop  
The development of Section 5 relied on two important 
propositions, concerned with the approximate bias and
variance of the method's implicit $\hatt R_{2m}$ estimator 
of the fourth derivative of $g$ at zero. 
The proofs of the propositions are given here. 

\subsection
{\sl A.1. Proof of the bias proposition.} 
We must work further with expression (5.4). 
The intention is to let $\hhh$ be at least moderately small, 
and look for expansions in terms $\hhh^2$, $\hhh^4$ and so on. 
The starting point is $\alpha_{2j}$ of (4.3), which we rewrite as 
$$\alpha_{2j}=\int H_{2j}(u)\exp(-\half u^2)G(\hhh u)\,\d u, $$
with $G$ being the smooth and symmetric function given in the proposition;
note that normality of $g$ is the same as saying that 
$G$ is identical to the constant $(2\pi)^{-1/2}$. 
This may be expanded via Taylor expansion for $G(\hhh u)$, 
for small $\hhh$, provided only that $g$ has the required 
number of derivatives. 
We find 
$${\alpha_{2j}\over (2\pi)^{1/2}}
=\sum_{i\ge0}{1\over (2i)!}\lambda_{2j,2i}G^{(2i)}(0)\,\hhh^{2i}
=\sum_{i\ge j}{1\over (2i)!}\lambda_{2j,2i}G^{(2i)}(0)\,\hhh^{2i}, $$
writing $\lambda_{2j,2i}$ for $\int H_{2j}(u)\phi(u)u^{2i}\,\d u$. 
Note that these vanish for $j>i$ by orthogonality of the Hermite polynomials. 
Hence the sum appearing in the second expression of (5.4) 
can be divided into three parts, like Gallia; 
$$\eqalign{
\sum_{j=0}^m&\sum_{i\ge j}{\lambda_{2j,2i}\over (2i)!}G^{(2i)}(0)
        {(-1)^j\over 2^j j!}\hhh^{2i}
        +\sum_{j=1}^m\sum_{i\ge j}{\lambda_{2j,2i}\over (2i)!}G^{(2i)}(0)
        {(-1)^j4j\over 2^j j!}\hhh^{2i-2} \cr 
&\qquad\qquad 
+\sum_{j=2}^m\sum_{i\ge j}{\lambda_{2j,2i}\over (2i)!}G^{(2i)}(0)
        {(-1)^j(4/3)j(j-1)\over 2^j j!}\hhh^{2i-4}. \cr}$$
Rearranging terms suitably gives 
$$\eqalign{
\sum_{i\ge0}&{1\over (2i)!}\Bigl\{\sum_{0\le j\le \min(i,m)}
        (-1)^j{\lambda_{2j,2i}\over 2^j j!}\Bigr\}
        G^{(2i)}(0)\,\hhh^{2i} \cr 
&\qquad\qquad 
+\sum_{i\ge0}{1\over (2i+2)!}\Bigl\{\sum_{0\le j\le \min(i,m-1)}
        (-1)^{j-1}2{\lambda_{2j+2,2i+2}\over 2^j j!}\Bigr\}
        G^{(2i+2)}(0)\,\hhh^{2i} \cr 
&\qquad\qquad
+\sum_{i\ge0}{1\over (2i+4)!}\Bigl\{\sum_{0\le j\le \min(i,m-2)}
        (-1)^j{1\over 3}{\lambda_{2j+4,2i+4}\over 2^j j!}\Bigr\}
        G^{(2i+4)}(0)\,\hhh^{2i} \cr
&\qquad\qquad
=A_0+A_2\hhh^2+A_4\hhh^4+A_6\hhh^6+\cdots, \cr}$$
say. 

\font\smalla=cmr9
\font\smallb=cmr8
\font\smallc=cmr7
\font\smalld=cmr6
All this simplifies to a spectacular degree. 
The leading constant term is found to be 
$G(0)-2G''(0)+(1/3)G^{(4)}(0)$ 
(assuming that at least $\alpha_0,\alpha_2,\alpha_4$ 
are included in the expansion), and by some calculations 
this is seen to be the same as $(1/3)g^{(4)}(0)(\sqrt{2}\sigma)^5$. 
The pleasant and surprising simplification is that all other 
terms simply v{\smalla a}{\smallb n}{\smallc i}{\smalld sh}, 
if only the size of $m$ permits full sums 
$0\le j\le i$ to be taken. That is, 
$$\sum_{0\le j\le i}(-1)^j{\lambda_{2j,2i}\over 2^j j!}=0
        \quad {\rm for\ }i=1,2,\ldots, \eqno({\rm A.1})$$
and the same thing happens for the two other sums involved. 
Proving this involves rather long calculations,
where a formula for $\lambda_{2j,2i}$ is necessary.
One such exploits the fact that 
$H_{2j}(x)=\sum_{l=0}^j(-1)^l w(2j,l)x^{2j-2l}$,
where $w(2j,l)=(2j)!/\{l!(2j-2l)!2^l\}$, 
see Fenstad and Hjort (1997); 
an alternative route starts with re-expressing 
$\alpha_{2j}/(2\pi)^{1/2}$ as 
$\int \phi(u)G^{(2j)}(\hhh u)\,\d u\,\hhh^{2j}$, 
after $2j$ partial integrations. These facts can be combined with 
$\E N^{2k}=(2k)!/(k!2^k)$ for standard normal even moments to give  
$$\lambda_{2j,2i}=\sum_{l=0}^j (-1)^l{(2j)!\over l!(2j-2l)!2^l}
        {(2j-2l+2i)!\over (j-l+i)!2^{j-l+i}} \quad {\rm for\ }j\le i. $$
This can be seen to lead to (A.1) by a suitable laborious induction proof. 

To demonstrate how this leads to the desired conclusion,
let us illustrate the induction step for $m=4$, 
so that the expansion includes coefficients of order 0, 2, 4, 6, 8,
and assume the previous steps $m=1,2,3$ have been taken care of.
Then $R_{2m}$ is equal to $R(f'')$ plus certain 
bias terms of the small order $\hhh^6$ 
(the $\hhh^2$ and $\hhh^4$ terms have already disappeared
by induction hypothesis). These terms are found by 
examining the general calculations above; 
$$\eqalign{&{1\over 6!}\sum_{0\le j\le 3}(-1)^j{\lambda_{2j,6}\over 2^j j!}
        G^{(6)}(0)\,\hhh^6
+{1\over 8!}\sum_{0\le j\le 3}(-1)^{j-1}2{\lambda_{2j+2,8}\over 2^j j!}
        G^{(8)}(0)\,\hhh^6 \cr 
&\qquad\qquad 
+{1\over 10!}\sum_{0\le j\le 2}(-1)^j{1\over 3}
        {\lambda_{2j+4,10}\over 2^j j!}G^{(10)}(0)\,\hhh^6. \cr}$$
The two first sums are zero and the last would also have been
had the sum been all the way to $j=3$. This leads to
the conclusion, via $\lambda_{10,10}=10!$, 
and ends our demonstration. 

\subsection 
{\sl A.2. Proof of the variance proposition.} 
By the initial rewriting of $\hatt R_{2m}$ in Section 5.3,
the problem of finding a useful expression for its variance 
may be attacked using tools of $U$-statistics. One has 
$$\Var\,\hatt R_{2m}={4\over n}{n-2\over n-1}\,{\rm cov}_{\rm be}
        +{2\over n(n-1)}\,{\rm cov}_{\rm wi}, \eqno({\rm A.2})$$
featuring between- and within-covariance terms. 
We tend to each in turn, and start with the between-term. Here 
$$\eqalign{{\rm cov}_{\rm be}
&={9\over (\sqrt{2}\sigma)^{10}}{\rm cov}\Bigl\{
\sum_{j=0}^m \hhh^{-1}c(j)p_{2j}\Bigl({\hhh^{-1}Y_{1,2}
        \over \sqrt{2}\sigma}\Bigr),
\sum_{l=0}^m \hhh^{-1}c(l)p_{2l}\Bigl({\hhh^{-1}Y_{1,3}
        \over \sqrt{2}\sigma}\Bigr)\Bigr\} \cr
&={9\over \tau^{10}}\sum_{j=0}^m\sum_{l=0}^m \Bigl[c(j)c(l)
   \int\int H_{2j}(u_1)\phi(u_1)H_{2l}(u_2)\phi(u_2) \cr  
&\qquad\qquad\qquad\qquad\qquad\qquad\qquad  
   \bar G(\hhh u_1,\hhh u_2)\,\d u_1\,\d u_2\Bigr] - R_{2m}^2, \cr}$$
in which $\tau$ is short hand for $\sqrt{2}\sigma$ and 
$$\bar G(x_1,x_2)=\exp(\half x_1^2+\half x_2^2)
        \bar g(\tau x_1,\tau x_2)\tau^2, $$ 
defined in terms of the density 
$$\bar g(y_1,y_2)=\int f(x+y_1)f(x+y_2)f(x)\,\d x $$
for a related pair of differences, 
say $(Y_{1,2},Y_{1,3})$, or $(X_2-X_1,X_3-X_1)$. 
To progress further, employ $2j$ partial integrations in the $u_1$ direction 
and $2l$ similar operations in the $u_2$ direction.
The double integral can then be expressed as 
$$\E_N\bar G_{2j,2l}(\hhh U_1,\hhh U_2)\,\hhh^{2j+2l}
        =\bar G_{2j,2l}(0,0)\,\hhh^{2j+2l}\{1+O(\hhh^2)\}, $$
in which $\bar G_{2j,2l}(x_1,x_2)$ is found by taking $2j$ 
derivatives in $x_1$ and $2l$ derivatives in $x_2$, and 
where the expectation is with respect to $U_1$ and $U_2$ 
being independent and standard normal. 

This can be employed in the double sum above.
Somewhat careful analysis is called for since the $c(j)$s
themselves have $1/\hhh^2$ and $1/\hhh^4$ terms,
and it becomes an accountant's challenge to keep track of 
all contributions. One has $c(0)=1$, 
$c(1)\hhh^2=-2-\half\hhh^2$,
$c(2)\hhh^4={1\over 3}+\hhh^2+{1\over 8}\hhh^4$, 
and we may put this to use in 
$$\sum_{k\ge 0}\sum_{i=0}^k c(k-i)c(i)\,
        \E_N\bar G_{2k-2i,2i}(\hhh U_1,\hhh U_2)\,\hhh^{2k}. $$
Terms up to order $k=4$ here, or order $\hhh^8$, 
must be monitored in order to chase the leading constant. 
After some analysis the result is of the form 
$T_0+\cdots+T_4$ plus $O(\hhh^2)$ or smaller terms, where 
the five $T_k$ terms involve various partial derivatives
of the $\bar G$ function at $(0,0)$. 
These in turn must involve the derivatives of the 
difference pair density $\bar g$ at zero, that is, the quantities 
$$R_{a,b}=\bar g_{a,b}(0,0)=\int f^{(a)}f^{(b)}f\,\d x. $$
One finds 
$$\eqalign{
T_0&=\bar G_{0,0}=R_{0,0}\tau^2, \cr
T_1&=-2(\bar G_{2,0}+\bar G_{0,2})=-4(R_{0,0}\tau^2+R_{0,2}\tau^4), \cr
T_2&={1\over 3}(\bar G_{4,0}+\bar G_{0,4})+4\bar G_{2,2}
        ={2\over 3}(3R_{0,0}\tau^2+6R_{2,0}\tau^4+R_{4,0}\tau^6) \cr
&\qquad\qquad
        +4(R_{0,0}\tau^2+2R_{0,2}\tau^4+R_{2,2}\tau^6), \cr
T_3&=-2(\bar G_{4,2}+\bar G_{2,4})
        =-{4\over 3}(3R_{0,0}\tau^2+9R_{0,2}\tau^4+6R_{2,2}\tau^6
        +R_{0,4}\tau^6+R_{2,4}\tau^8), \cr
T_4&={1\over 9}\bar G_{4,4}
        ={1\over 9}(9R_{0,0}+36R_{0,2}\tau^4+6R_{0,4}\tau^6
        +36R_{2,2}\tau^6+12R_{2,4}\tau^8+R_{4,4}\tau^{10}). \cr}$$
And adding these most terms make their polite excuses and 
one is left with simply $(1/9)R_{4,4}\tau^{10}$. 
Thus the remaining leading term of the first part of the 
(A.2) variance is simply $(4/n)(R_{4,4}-R_{2m}^2)$ 
plus various $\hhh^2/n$ terms. 

The second part of the (A.2) variance is easier to deal with.
One finds that ${\rm cov}_{\rm wi}$ can be written
$${9\over \tau^{10}}{1\over \hhh}\int
        \Bigl\{\sum_{j=0}^m c(j)p_{2j}(u)\Bigr\}^2
        g(\hhh\tau u)\tau\,\d u-R_{2m}^2, $$
and this is of size $O(1/\hhh^9)$ since $c(j)$s are of size 
$1/\hhh^4$ for $j\ge2$. 

\bigskip
{\bf Acknowledgements.} 
This work started out during a pleasant visit to 
the Institute of Statistics at the Universit\'e Catholique de 
Louvain-la-Neuve. My work related to this report 
has benefitted from and will continue to be influenced 
by stimulating conversations, in vivo and in electro, 
with Ir\`ene Gijbels and Chris Jones. 

\bigskip
\centerline{\bf References}

\def\ref#1{{\noindent\hangafter=1\hangindent=20pt
  #1\smallskip}}          
\parindent0pt
\baselineskip11pt
\parskip3pt 
\medskip 

\ref{%
Bickel, P.J.~and Ritov, Y. (1988).
Estimating integrated squared density derivatives.
{\sl Sankhy\=a Series A} {\bf 50}, 381--393.} 

\ref{%
Bowman, A. (1984).
An alternative method of cross-validation for the smoothing
of density estimates.
{\sl Biometrika} {\bf 71}, 353--360.}

\ref{%
Byholt, M.~and Hjort, N.L. (1999).
Sometimes nonparametrics beat parametrics,
even when the model is right.
{\sl American Statistician}, to appear.} 

\ref{%
Chiu, S.-T. (1996). 
A comparative review of bandwidth selection 
for kernel density estimation.
{\sl Statistica Sinica} {\bf 6}, 129--145.} 

\ref{%
Fan, J.~and Marron, S.J. (1992).
Best possible constant for bandwidth selection. 
{\sl Annals of Statistics} {\bf 20}, 2057--2070.}

\ref{%
Fenstad, G.U.~and Hjort, N.L. (1997).
Two Hermite expansion estimators, 
and a comparison with the kernel estimator.
Unpublished manuscript.} 

\ref{%
Hall, P.~and Marron, J.S. (1987).
Estimation of integrated squared density derivatives.
{\sl Statistics and Probability Letters} {\bf 6}, 109--115.}

\ref{%
Hall, P., Marron, J.S.~and Park, B.U. (1992).
Smoothed cross validation. 
{\sl Probability Theory and Related Fields} {\bf 92}, 1--20.} 

\ref{%
Hall, P., Sheather, S.J., Jones, M.C.~and Marron, J.S. (1991).
On optimal data-based bandwidth selection in kernel density estimation.
{\sl Biometrika} {\bf 78}, 263--269.} 

\ref{%
Hjort, N.L. (1993).
{\sl Density Estimation and Smoothing.} 
Lecture Notes compendium, Department of Mathematics, University of Oslo.} 

\ref{%
Hjort, N.L.~and Glad, I.K. (1995).
Nonparametric density estimation with a parametric start.
{\sl Annals of Statistics}. }

\ref{%
Hjort, N.L.~and Jones, M.C. (1995).
Better rules of thumb for choosing bandwidth in density estimation.
Unpublished manuscript.} 

\ref{%
Hjort, N.L.~and Jones, M.C. (1996).
Locally parametric nonparametric density estimation.
{\sl Annals of Statistics} {\bf 24}, 1619--1647.} 

\ref{%
Jones, M.C. (1991).
The roles of ISE and MISE in density estimation.
{\sl Statistics and Probability Letters} {\bf 12}, 51--56.}

\ref{%
Jones, M.C., Marron, J.S.~and Sheather, S.J. (1996).
Progress in data-based bandwidth selection for kernel density estimation.
{\sl Computational Statistics} {\bf 11}, 337--381.} 

\ref{%
Jones, M.C.~and Sheather, S.J. (1991).
Using non-stochastic terms to advantage 
in kernel-based estimation of integrated squared density derivatives.
{\sl Statistics and Probability Letters} {\bf 11}, 511--514.}

\ref{%
Kim, W.C., Park, B.U.~and Marron, J.S. (1994).
Asymptotically best bandwidth selectors in kernel
density estimation.
{\sl Statistics \& Probability Letters} {\bf 19}, 119--127.}

\ref{%
Loader, C. (1996). 
Local likelihood density estimation.
{\sl Annals of Statistics} {\bf 24}, 1602--1618.} 

\ref{%
Marron, S.~and Wand, M.P. (1992).
Exact mean integrated squared error.
{\sl Annals of Statistics} {\bf 20}, 712--736.}

\ref{%
M\"uller, H.-G. (1985).
Empirical bandwidth choice for nonparametric kernel regressions 
by means of pilot estimators.
{\sl Statistics and Decisions}, supplement {\bf 2}, 193--206.} 

\ref{%
Rudemo, M. (1982).
Empirical choice of histograms and kernel density estimators.
{\sl Scandinavian Journal of Statistics} {\bf 9}, 65--78.} 

\ref{%
Scott, D.W. (1992).
{\sl Multivariate Density Estimation:
Theory, Practice, and Visualization.}
Wiley, New York.}

\ref{%
Sheather, S.J.~and Jones, M.C. (1991).
A reliable data-based bandwidth selection method for kernel
density estimation.
{\sl Journal of the Royal Statistical Society} {\bf B 53}, 683--690.} 

\ref{%
Silverman, B.W. (1986).
{\sl Density Estimation for Statistics and Data Analysis.}
Chapman and Hall, London.}

\ref{%
Simonoff, J.S. (1996).
{\sl Smoothing Methods in Statistics.}
Springer-Verlag, New York.} 

\ref{%
Staniswalis, J. (1989).
Local bandwidth selection for kernel estimates. 
{\sl Journal of the American Statistical Association} {\bf 84}, 284--288.} 

\ref{%
Stone, C.J. (1984).
An asymptotically optimal window selection rule
for kernel density estimation.
{\sl Annals of Statistics} {\bf 12}, 1285--1297.}

\ref{%
Wand, M.P.~and Jones, M.C. (1995).
{\sl Kernel Smoothing.}
Chapman \& Hall, London.}

\bye